\documentclass[aps,twocolumn,showpacs,showkeys,amsmath,amssymb,floatfix,superscriptaddress,pre]{revtex4-1}
\usepackage[utf8]{inputenc}
\usepackage{textcomp}
\usepackage{graphicx}
\usepackage{mathcomp}
\usepackage{color}

\newcommand{\be}{\begin{equation}}
\newcommand{\ee}{\end{equation}}

\newcommand{\bea}{\begin{eqnarray}}
\newcommand{\eea}{\end{eqnarray}}

\begin{document}

\title{Residual entropy and waterlike anomalies in the repulsive one dimensional lattice gas}
\author{Fernando Barbosa V. da Silva}
\affiliation{Instituto de F\'isica, Universidade de Bras\'ilia, Bras\'ilia-DF, Brazil}
\affiliation{Instituto Federal de Bras\'ilia, Campus S\~ao Sebasti\~ao, S\~ao Sebasti\~ao-DF, Brazil}
\author{Fernando Albuquerque Oliveira}
\email{fao@fis.unb.br}
\affiliation{Instituto de F\'isica, Universidade de Bras\'ilia, Bras\'ilia-DF, Brazil}
\author{Marco Aur\'elio A. Barbosa}
\thanks{Corresponding Author}
\email{aureliobarbosa@gmail.com}
\affiliation{Programa de P\'os-Gradua\c{c}\~ao em Ci\^encia de Materiais, Faculdade UnB Planaltina, Universidade de Bras\'ilia, Planaltina-DF, Brazil}

\date{\today}

\begin{abstract}
The thermodynamic and kinetics of the one dimensional lattice gas with repulsive interaction is investigated using transfer matrix technique and Monte Carlo simulations. This simple model is shown to exhibit waterlike anomalies in density, thermal expansion coefficient and self diffusion. An unified description for the thermodynamic anomalies in this model is achieved based on the ground state residual entropy which appears in the model due to mixing entropy in a ground state  phase transition. 
\end{abstract}


\maketitle

\section{Introduction}

Water is special fluid for its biological relevance and technological applications but most intriguing is that it presents thermodynamic and dynamic properties properties with anomalous (or unusual) behavior~\cite{Franks:water:matrix}. The origin of its anomalous properties is actively discussed in the literature, with different thermodynamic scenarios competing to describe its behavior on regular and metastable regimes~\cite{debenedetti03:review, malenkov2009:jpcm:review}. Among alternative views on water thermodynamics it should be relevant to mention the second critical point hypothesis~\cite{poole92:nat} and the singularity free scenario~\cite{sastry96:pre,sastry98:jcp}, which will be relevant in the context of the current work.

The second critical point  hypothesis was first proposed by Poole {\it et al.} based on computer simulations of atomistically detailed water models, and relates the observed divergence on water's thermodynamic response functions to the critical end point of a deeply metastable liquid-liquid phase transition~\cite{poole92:nat}. Singularity free scenario was proposed by Sastry~\textit{et al.} using lattice models for liquid water which supported the idea that waterlike  anomalies indeed exist, at least in lattice systems, without a liquid-liquid coexistence~\cite{sastry96:pre}. In some simplified models of fluid it was possible to obtain both second critical point and singularity free scenario by adjusting some physical parameters of the system~\cite{giancarlo03:pre,heckmann2013:jcp}
For instance, in the model proposed by Franzese~\textit{et al.}~\cite{giancarlo03:pre} it is possible to observe the liquid-liquid coexistence to shrink and the temperature of second critical point  decrease to $T_c \rightarrow 0$ by decreasing the correlation between hydrogen bonds inside within individual molecules.

Lattice models of fluid have been extensively used to investigate the above mentioned anomalous properties of water due to possibility of obtaining analytical or numerical results, while exploring a wide range of physical parameters. In this direction, both thermodynamics~\cite{barbosa08,marcia05:jcp, giancarlo03:pre, sastry96:pre,sastry98:jcp, heckmann2013:jcp} and kinetics~\cite{szortyka10:jcp, giancarlo08:prl} where investigated in lattice models with waterlike behavior. 
Nevertheless, approximations employed in two and three dimensions(3D), and even some exact solutions in one dimension (1D), tend to generate complex sets of equations whose analyses is often performed numerically. Thus, it should desired to design models for which one could obtain simple analytical expressions connecting thermodynamic anomalous behavior to phase transitions and critical behavior. 

To achieved this goal we previously investigate 1D lattice models with pair interaction between the first neighboring molecules, with interactions spanning two~\cite{barbosa11:jcp} and three lattice sites~\cite{barbosa13:pre}. While in Ref.~\cite{barbosa11:jcp} both van der Waals and hydrogen bond like interactions were used, resulting in a line of temperature of maximum density associated to a ground state phase transition (GSPT), in Ref.~\cite{barbosa13:pre} it was proposed a core-softened fluid with pair interactions up to three sites, resulting in two temperature of maximum density lines associated to two GSPT. Besides obtaining exact results, in the latter work we used an analytical approximation in the neighborhood of the critical point to obtain a simple expression for Gibbs free energy, and used it to mathematically study the relation between anomalous density behavior and GSPT.

In this work we proceed on this direction by investigating the repulsive 1D lattice gas, which is even simpler than our previous models and presents waterlike anomalies in density, thermodynamic response functions and self diffusion constant. The model was studied through transfer matrix technique, the Takahashi method (within a two state approximation, as will be discussed latter), and Monte Carlo simulations. With the results obtained from these techniques a connection between temperature of maximum density and GSTP was found as in a previous work with more complex models~\cite{barbosa13:pre}. In addition, it was also found that GSPT does present a residual entropy, due to phase mixing, and it is shown that this property is fundamental in determining waterlike anomalies for the model considered here. Finally, a comparison between regions with density and diffusion anomaly indicated that this  model presents so called \textit{hierarchy of anomalies}~\cite{debenedetti01:nat}.

\section{\label{thermo}Thermodynamics}

\subsection{Model and Ground State}

By defining the presence of a particle on site $k$ with an occupation variable $\eta_k = 1$ (and absence with $\eta_k=0$) the effective Hamiltonian of the 1D repulsive lattice gas in the grand canonical ensemble becomes:
\be
\label{eq:H}
\mathcal{H} = \sum_i \epsilon{}{\eta{}}_i{\eta{}}_{i+1}-\sum_i \mu{}{\eta{}}_i,
\ee 
where $\epsilon>0$ is the strength of the (repulsive) interaction and $\mu$ is the chemical potential. Note that this Hamiltonian can be seen as a special case from our previous model for liquid water~\cite{barbosa11:jcp} by setting $\epsilon_{hb}=0$ and $\epsilon_{vdw}=-\epsilon$.

The ground state of the unidimensional repulsive lattice gas is composed by three phases: a gas in which the lattice is empty, a dense fluid (DF), presenting  a completely filled lattice, and a half-filled lattice with each particle separated by holes, here denominated as 
softened fluid (SF)\footnote{Notation for this phase has changed from bonded fluid to softened fluid, since our previous work, because there are no~\textit{hydrogen bonded} in the current context.}. In a lattice with $L$ sites separated by a distance $l$, periodic boundary conditions, and $N$ particles, the enthalpy per particle of these phases are $h_\texttt{gas}=0$, $h_{\texttt{DF}}=\epsilon + Pl$ and $h_{\texttt{SF}}=2Pl$, with $P$ meaning pressure. Considering this, the SF is stable in the ground state for pressures $0 < P \leq P_c$ with $P_c l = \epsilon$. 

Two ground state phase transitions are present at null temperature and were calculated by equating the enthalpy per particle at different phases: a G-SF at $P=0$ and a SF-DF transition at $P_c$, which will be called ``second critical point'' as in previous works~\cite{stanley99:pre,barbosa11:jcp,barbosa13:pre}.

\subsection{Transfer Matrix Technique}
Thermodynamics of the model proposed here is obtained using transfer matrix technique and will be described shortly. Other derivations using this technique can be found elsewhere~\cite{barbosa10:jcp}. We start from the grand canonical partition function as a trace of a matrix
\begin{equation}
\Xi{}\left(T,L,\mu\right)= \sum_{\vec \eta} e^{-\beta \mathcal{H} } =  Tr \left \{ \mathcal{P}^L  \right \},
\label{equ:traco}
\end{equation}
where $\beta= 1/k_BT$, $\vec \eta = \{ \eta_1, \ldots, \eta_N  \}$, and the elements of $\mathcal{P}$ are given by
\begin{equation}
\mathcal{P}_{\eta \eta' }=e^{-\beta{}\left(\epsilon{}{\eta{}}{\eta{}}'-\
\mu{}{\eta{}}\right)}.
\label{eq:matriz}
\end{equation}
In the thermodynamic limit only the largest eigenvalue of $\mathcal{P}$ will contribute to the partition function:
\begin{equation}
\Xi{}\left(T,L,\mu\right)={\lambda{}}^L=e^{\beta{}P l L}.
\end{equation}
The eigenvalue $\lambda$ is related to the fugacity $z=e^{\beta \mu}$ via characteristic function of $\mathcal{P}$,
\begin{equation}
\left(az-\lambda{}\right)\left(1-\lambda{}\right)-z=0, \label{eq:z}
\end{equation}
with $a = e^{-\beta \epsilon}$.

Next we introduce reduced variables $t=k_BT/\epsilon$, $p=Pl/\epsilon$, $g = \mu / \epsilon$,  $v = V/Nl$ and $s = S/{Nk_B}$. Since the equation of state~(\ref{eq:z}) connects the Gibbs free energy $g$ to thermodynamic variables $t$ and $p$, it is possible to obtain the `volume' $v$ and entropy $s$ per particle as
\be 
 v = \left(\frac{\partial g}{\partial p}\right)_p, \hspace{0.4cm} s = -{\left(\frac{\partial g}{\partial t}\right)}_p. \nonumber
\ee

With these definitions, density can be calculated as
\be 
\label{eq:density}
\rho  =  \frac{1}{v} = \frac{\left(1-\lambda{}\right)[a\left(1-\lambda{}\right)-1]}{a{(1-\lambda{})}^2-1+2\lambda{}},
\ee 
and entropy becomes
\be
\label{eq:entropy}
s =  -\ln{\left(z\right)}+\frac{p}{t} + \frac{za[\left(1-\lambda\right)+p\lambda]-p \lambda^2}{zt \left [a\left(1-\lambda\right)-1 \right ]} .
\ee
Thermodynamic response functions, such as thermal expansion coefficient $\alpha$, isothermal compressibility $k_T$, and isobaric heat capacity $c_P$ are calculated from standard definitions~\cite{salinas:introduction} which, within the convention adopted here, reads
\begin{subequations}
\bea
\alpha v & = & \left ( \frac{\partial v}{\partial t} \right )_p,\label{eq:alphav}\\
k_t v & = & -\left ( \frac{\partial v}{\partial p} \right )_t\\
c_p & = & t\left ( \frac{\partial s}{\partial t} \right )_p. \label{eq:cp}
\eea
\end{subequations}
Explicit (and exact) expressions for these functions are too lengthy to be reproduced here. The temperature of maximum density line was calculated from~(\ref{eq:alphav}) by numerically solving $\alpha v=0$.

\begin{figure}
\begin{center}
\includegraphics[scale=0.7]{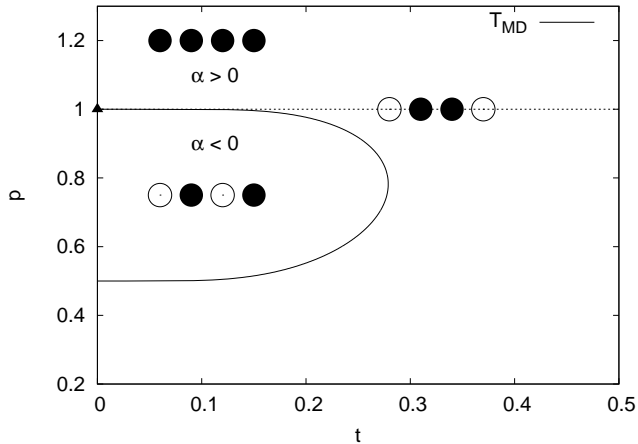} 
\caption{Location of temperature of maximum density line ($T_{MD}$) in the $p$ vs. $t$ phase diagram. The ground state phase transition between the softened fluid and dense fluid is indicated with a triangle, while its pressure value is marked with a dotted line. An example of the mixed SF/DF state is shown along the dotted line, used to indicate the critical pressure $p=p_c=1$.}
\label{fig:diagram}
\end{center}
\end{figure}

\begin{figure}
\begin{center}
\includegraphics[scale=0.7]{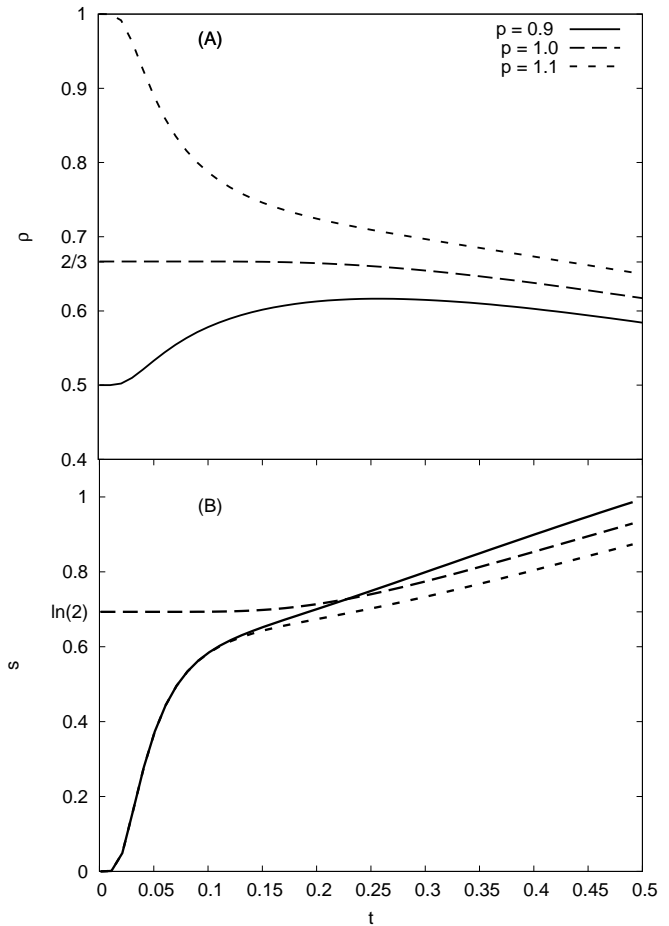}
\caption{The temperature dependence of (a) density and (b) entropy for pressures equal, slightly below and above the ground state phase transition at $p=p_c=1$.}
\label{fig:dens}
\end{center}
\end{figure}

\subsection{\label{two-state}Two states approximation}

To obtain further theoretical insight about this system, it is interesting to investigate its entropy in the neighborhood of the second critical point using a two state approximation based on a lattice version of the Takahashi method, as employed in Ref.~\cite{barbosa13:pre}.  In that work the Gibbs free energy per particle was approximated near the GSPT by realizing that on this region the probability of finding local states different from those of the coexisting structures was vanishing small. In the current model these states would correspond to softened and dense fluids, whose ground state local molecular spacing are $r_S=2r_D=2l$. Thus, the Gibbs free energy (non-reduced units) becomes approximately:
\begin{equation}
G/N \approx h_c -k_B T \ln \left [ e^{-\beta (P-P_c)r_S} + e^{-\beta (P-P_c)r_D}\right ],
\label{eq:g-pc}
\end{equation}
where $h_c$  is a constant identical to the local microscopic enthalpy at the GSPT (see Ref.~\cite{barbosa13:pre}). Within the notation adopted here,  the Gibbs free energy on Eq.~(\ref{eq:g-pc}) reads
\be
 g = \frac{h_c}{\epsilon} +  \frac{3}{2}\Delta p  - t \ln \left [  2 \cosh \left ( \frac{\Delta p}{2t} \right )  \right ],
\label{eq:g-two-states}
\ee
with $\Delta p = p-p_c$. From the latter, molecular volume and entropy becomes
\be
v = \frac{3}{2}-\frac{1}{2}\tanh \left( \frac{\Delta p}{2t} \right),
\label{eq:v-2s}
\ee
and
\be
s = \ln \left [ 2 \cosh \left( \frac{\Delta p}{2t} \right) \right ] - \left ( \frac{\Delta p}{2t} \right) \tanh \left ( \frac{\Delta p}{2t} \right). 
\label{eq:entropy-2s}
\ee 
It is possible to see that by approaching the GSPT at constant pressure, i.e., $(t \rightarrow 0, p = p_c = 1 )$, one obtains the residual entropy $s \rightarrow \ln 2$, discussed before. For any other route used to approach the ground state, entropy results in a null value, indicating that in this model residual entropy only occurs exactly at the GSPT. Molecular volume also depends on the path along which criticality is approached, resulting in
\begin{subequations}
\bea
\lim_{p \rightarrow p_c }  \lim_{t \rightarrow 0} v & = &  \left \{
\begin{array}{lc}
2, & \qquad  p \rightarrow p_c^- \\
1, & \qquad  p \rightarrow p_c^+ \label{eq:v-lowt-1}
\end{array}
 \right . \\
\lim_{t \rightarrow 0 }  \lim_{p \rightarrow p_c} v & = &  \frac{3}{2}. \label{eq:v-lowt-2}
\eea 
\end{subequations}
It is important to investigate the critical behavior of the system near the GSPT, by calculating analytical expressions for thermodynamic response functions~(\ref{eq:alphav})-(\ref{eq:cp}), within the two-states approximation adopted here. It is convenient to define $f(x)=1-\tanh^2x$, with $x=\Delta p/2t$, such that
\begin{subequations}
\bea
\alpha v & = & \Delta p^{-1} x^2 f(x) ,\label{eq:alphav-2s}\\
k_t v & = & \frac{f(x)}{4t},\label{eq:ktv-2s}\\
c_p & = &   x^2f \left (x \right) . \label{eq:cp-2s}
\eea
\end{subequations}
Since  $0 \leq f(x) \leq 1$, thermodynamic response functions diverge toward criticality as $k_t \sim t^{-1}$, $\alpha~\sim  {\Delta p}^{-1}  x^2$ and $c_p \sim x^2$. The low temperature behavior on response functions is consistent with waterlike behavior and the critical exponents for $k_t$ and $c_p$ are identical to those found in a continuous one dimensional core softened model proposed by Sadr-Lahijany {\emph et al}~\cite{stanley99:pre} in the same context.

\subsection{Fluid Structure}

\begin{figure}
\begin{center}
\includegraphics[scale=0.7]{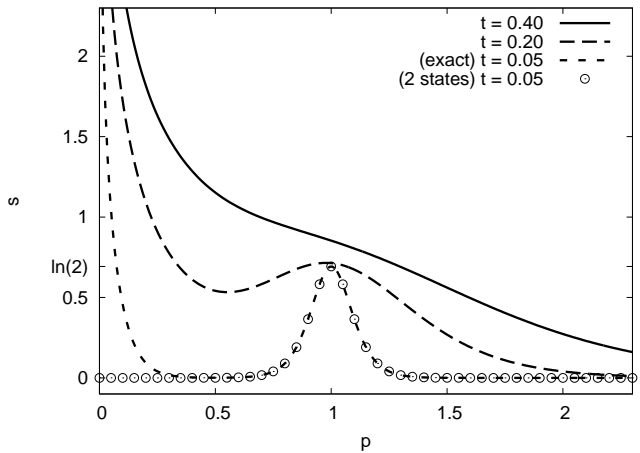}
\caption{Entropy as a function of pressure at a fixed temperatures from exact (continuous and dashed lines) and two state approximation (circles). Note that entropy presents a maximum value at low temperatures.}
\label{fig:entro}
\end{center}
\end{figure}

\begin{figure}
\begin{center}
\includegraphics[scale=0.7]{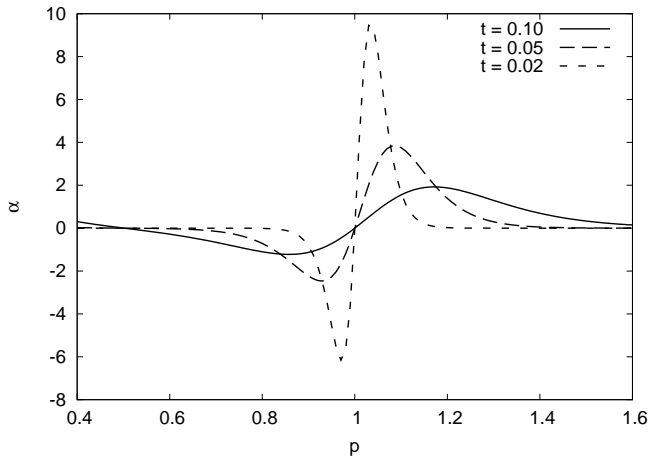}
\caption{Thermal expansion coefficient, $\alpha$, as a function of pressure at fixed temperatures. At low temperatures, an oscillatory behavior is observed in the neighborhood of the ground state critical pressure $p_c=1$.}
\label{fig:alpha}
\end{center}
\end{figure}

We start discussing our model's liquid structure from its exact pressure vs. temperature phase diagram, shown in Fig.~\ref{fig:diagram}.
The model presents a GSPT between softened and dense phases at $p=p_c=1$, from which a temperature of maximum density line, $T_{MD}$, emerges separating a region with density anomaly ($\alpha < 0$) from a region with normal density behavior ($\alpha>0$). This line was obtained by numerically solving the equation $\alpha=0$, and was found to extend up to a temperature $t=0.279(1)$, and proceed retracing to lower temperatures ending exactly at pressure $p_{low}=1/2$. An explanation for this precise number can be obtained is as follows: at null temperature, the entropy gain for contracting or expanding the softened fluid by a single site will be same (a lattice defect), thus the system contracts or expands through a competition between fluctuations in energy and volume, which mathematically corresponds to free energy fluctuations of $\delta g_{con} = \epsilon -Pl$ (contraction) or $\delta g_{exp} = +Pl$ (expansion). By equating both variations it follows that $p_{low}= 1/2$.

Another interesting feature of this phase diagram is the mixed SF/DF state, which occurs near the GSPT. This can be observed in Fig.~\ref{fig:dens}~(a), where density is shown as a function of temperature for the critical pressure $p_c=1$ and values slightly below and above $p_c$. Density anomaly is evident for $p<p_c$ while density behaves normally for $p>p_c$. Exactly at $p=p_c=1$ density is kept constant up to $t\approx 0.15$. Density behaves this way because $\alpha \propto (\partial \rho / \partial T )_P \approx 0$ near the $T_{MD}$ line.

The entropy increase with temperature is depicted in Fig.~\ref{fig:dens}~(b), for the pressures investigated in Fig.~\ref{fig:dens}~(a). From this data it is evident that a residual entropy occurs exactly at $p=p_c=1$, while for other pressures ground state entropy start to increase from a null value. It is well known that, due to the Maxwell relation
\be
- \left ( \frac{\partial S}{\partial P} \right)_{T,N} = \left ( \frac{\partial V}{\partial T} \right)_{P,N},
\label{eq:Maxwell}
\ee
a density maximum at fixed pressure is mathematically equivalent to an entropy maximum as function of pressure at fixed temperature. Thus, in Fig.~\ref{fig:entro}, entropy is  investigated as a function of pressure (at fixed temperatures) using exact and approximate expressions obtained previously, Eqs.~(\ref{eq:entropy}) and~(\ref{eq:entropy-2s}). It is  important to note accuracy of the two states approximation near the transition. The relevance of a residual entropy for the appearance of a $T_{MD}$ line is evident since a region of anomalous entropy increase appears at low temperature in the neighborhood of the SF/DF transition. What is interesting in this behavior is that entropy anomaly appears naturally due to residual entropy, whose origin is the two phase mixture in the GSPT. In this way, anomalous entropy behavior (and anomalous density) is intrinsically associated to phase transition in our 1D model since mixing and critical behavior cannot be dissociated when continuous and discontinuous transitions are collapsed in the same point.

Residual entropy also explains the oscillatory behavior observed on thermal expansion coefficient ($\alpha$) as a function of pressure, near the critical transition, Fig.~\ref{fig:alpha}. This behavior was observed previously~\cite{barbosa11:jcp,barbosa13:pre} and its relation with entropy anomaly is simple: below the critical pressure $\alpha$ must be negative due to the fast increase in entropy towards its residual value $\ln2$, and above $p_c$ entropy must fast decrease to a low value, since it is null at $t=0$. 

It's worth mentioning that two states approximation also predicts that isothermal compressibility and constant pressure heat capacity increase while lowering temperature, at pressures near the critical one, through Eqs.~(\ref{eq:ktv-2s}) and~(\ref{eq:cp-2s}). Both response functions must present maximum values as a function of temperature, for pressures near the critical one. These results are not shown here but they can be rationalized as follows: the increase in isothermal compressibility should be expected near the GSPT because while increasing pressure in that region the molecular volume is abruptly changing from typically softened fluid to typically dense fluid. On the other hand, the existence of a maximum as a function of temperature in $c_p$, at pressures near the critical value, can be predicted from Eq.~(\ref{eq:cp}) and Fig.~\ref{fig:dens} (b). Its explanation proceeds as follows: let us consider a system in the softened fluid, at $t=0$ and $p<p_c$, where all pair of particles are separated by a hole. In this state, while increasing the temperature entropy will preferentially be increased by allowing neighboring particles to contract. As temperature increases further, an even higher number of particles pairs are allowed to fluctuate between softened and dense fluid states, but the nearest the system is to the critical pressure, the fastest is the approach to an entropic value of $\ln 2$, since the enthalpy difference between the two states is proportional to $\Delta p$. At higher temperatures this mixing mechanism will compete with translational diffusion, i.e., volume expansion, there must be a maximum in $(\partial s/\partial t)_p$ between those temperatures were each mechanism is dominating the entropic behavior of the system.

\section{\label{MC}Monte Carlo}

\begin{figure}
\begin{center}
\includegraphics[scale=0.7]{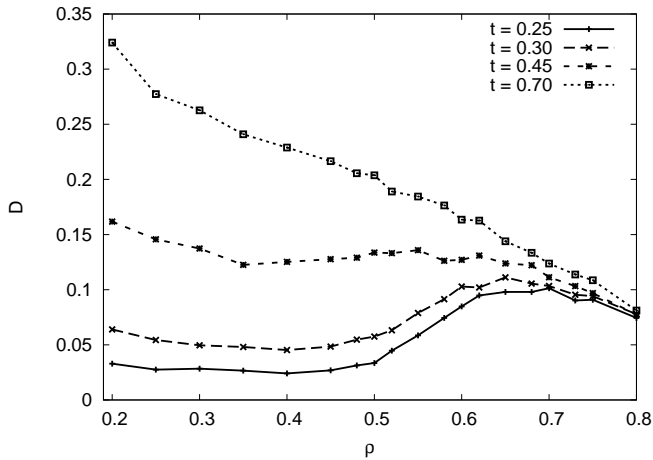}
\caption{Self-diffusion constant as a function of density at fixed temperature.}
\label{fig:difu}
\end{center}
\end{figure} 

\begin{figure}
\begin{center}
\includegraphics[scale=0.4]{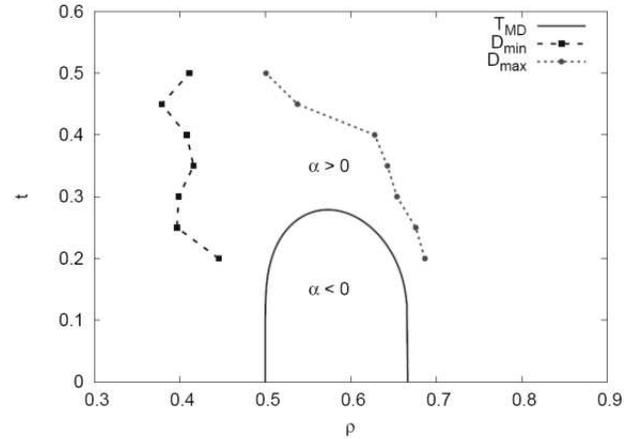}
\caption{The location of the maximum (circles) and minimum (square) values of the self-diffusion constant are 
compared to the line of temperature of maximum density (line) in the $t$ vs. $\rho$ phase diagram.}
\label{fig:hierarchy}
\end{center}
\end{figure}

Standard NVT Monte Carlo simulations with Metropolis algorithm were used to calculate the self diffusion constant of our model as a function of density and temperature~\cite{binder,comment2}. The system dynamics proceeds as follows: particles seated on site $k$ were allowed to jump to nearest neighbor ($k\pm 1$) and next nearest neighbor sites ($k\pm 2$). Jumping to next nearest neighbor sites is necessary for particles to diffuse in a 1D lattice, because there is no way for them to turn around their neighboring particles (the same strategy was used in our previous paper~\cite{barbosa11:jcp}. System size was equal to $L=10^3$ and simulation times were equal to or higher than $10^7$ Monte Carlo steps after an equilibrium time of $10^6$ Monte Carlo steps.

As in Refs.~\cite{barbosa11:jcp,marcia07:pa,marcia07:pa2}, self diffusion constant was obtained by linearly adjusting the mean square displacement of particles to the Einstein-Smoluchowski equation for the Brownian motion in one dimension,
\begin{equation}
\left \langle \frac {1}{N} \sum_{k=1}^{N} \left [ X_k(\tau) - X_k(0) \right ]^2  \right \rangle_\mathcal{R} = 2 D (t,\rho) \tau
\end{equation}
where $X_k(\tau)$ is the position of particle $k$ at time $\tau$ and $\mathcal{R} \geq 5$ is the number of different initial conditions used to average square displacements.

Fig.~\ref{fig:difu} exhibits $D(\rho,t)$ for different isotherms. At higher temperatures, as illustrated for $t=0.7$, the diffusion constant is `normal' and monotonically decreases with density. For $t < 0.55$ minimum and maximum diffusion appears, delimiting a region where diffusion anomalously increase with density, as observed experimentally~\cite{angell:jcp76}. 

In liquid water, density anomaly is associated to diffusion anomaly through a hierarchy of anomalies, with the former enclosing the latter in the phase diagram~\cite{debenedetti01:nat}. To investigate this hierarchy diffusion isotherms were adjusted to a fourth degree polynomial in the neighborhood of extremum temperatures. At low temperatures, the location of maximum and minimum diffusion were estimated using six points around each simulated extremum, and the best estimate for their values were taken from the analytically adjusted function. On intermediate temperatures, where maximum and minimum where too close, a single function and a higher number of points were used to adjust diffusion isotherms. The result is shown in Fig.~\ref{fig:hierarchy}, where is possible to see that an strict hierarchy of anomalies is found, with the same order observed in atomically detailed models of liquid water~\cite{debenedetti01:nat}.

\section{\label{final}Final discussions and conclusions}

We investigated the exact thermodynamics of the one dimensional repulsive lattice gas model using transfer matrix technique. In the ground state, the model presents a phase transition between a softened fluid and a dense fluid, which is characterized by a mixture between two states. The main consequence of this mixed state is a residual entropy in a single point in the pressure vs. temperature phase diagram. Using thermodynamic relations, it was argued that the presence of a residual entropy in a single point is consistent with a temperature of maximum density line ($\alpha=0$) emanating from the ground state phase transition. Considering this it is possible to state that ground state phase transition, residual entropy and density anomalies are thermodynamic properties of our model which are intrinsically related and cannot be dissociated from each other.

Further discussion about ground state phase transitions in one dimensional models is relevant at this point. First of all, the two transitions observed in our model can be seen as remanent from phase transitions on higher dimensional systems collapsing to $T \rightarrow 0$ as dimensionality is reduced to $d \rightarrow 1$. This collapse is also evident on the analytical solution of a lattice gas inside the Bethe lattice: in this case, while continuously reducing the coordination number the liquid-gas phase transition continuously shrinks to $T\rightarrow 0$~\cite{hu98:pre,barbosa10:jcp}. This is a relevant issue for one dimensional systems because the observed ground state phase transitions incorporate elements from both continuous and discontinuous phase transitions~\footnote{While crossing the neighborhood of the ground state phase transitions, in the $P$ vs. $T$ phase diagram, it is possible to observe the discontinuity in extensive variables (\textit{e.g.} volume) and the diverging behavior on response functions (\textit{e.g.} thermal expansion coefficient)}. While usually a discontinuous transition allows for any mixtures between two coexisting phases, when a continuous transition is collapsed on it, any mixture except for that maximizing the entropy is forbidden. Thus, for fluid-fluid transition entropy is maximized for equal distribution of randomly mixed interactions compatible with softened and dense fluids, resulting in an entropy per particle $s = S/N = k_b \ln 2$ (as in a coin toss game). 

The viability of using a simple two states approximation to accurately describe the system in the neighborhood of the ground state phase transition is another issue deserving attention, since there is a long tradition on using two state models to describe liquid water~\cite{malenkov2009:jpcm:review,sanches87,tanaka2014:nature}. Recently, a modified regular solution model within a two solvent approach was adjusted to computer simulation data from ST2 model, being accurately used to describe computational data above the second critical point, in a region called Widom line~\cite{poole11:prl}. This raises the possibility of connecting the current approach to more complex, higher dimensional and off-lattice models. Nevertheless, it should be emphasized that the two states character of the approximation employed here is conceptually distinct from the ones implemented in Ref.~\cite{poole11:prl}, where two states refers to two fluids with distinct free energies. Our approach resembles more the approximation implemented in Ref.~\cite{heckmann2013:jcp}, in the sense that two states are localized microscopic states with well defined local energies and entropies, which are useful for truncating the partition function in some kind of expansion. Certainly it should be interesting to link these different views, but this is outside the scope of the current work.

To finish, it should be mentioned that in a previous work on lattice models with waterlike behavior~\cite{barbosa11:jcp} we found an oscillatory behavior on thermal expansion coefficient $\alpha$ associated to density anomaly and ground state phase transition. More recently this effect was shown to be general, being a feature of any ground state phase transition between different fluid structures in one dimensional lattice models with interactions restricted to first neighboring particles and, inspired by these models, it was possible to design a three dimensional spherically symmetric pair potentials with two liquid-liquid phase transitions, both of them associated to temperature of maximum density lines~\cite{barbosa13:pre}. The oscillatory behavior on $\alpha$, as well as the anomalous behavior on isothermal compressibility and constant pressure heat capacity, can be explained in terms of a residual entropy using thermodynamical and statistical arguments, while for the current model the the relation between GSPT, residual entropies and waterlike behavior can be mathematically described within the two states approximation.

Obviously we are tempted to ask whether such entropic effects couldn't play an important rule in more complex systems, such as higher dimensional and off-lattice models. Current work is being done on this direction through molecular dynamics simulations of the 3D core softened models, as in Ref.~\cite{barbosa13:pre}.

This work has been supported by CNPq and FAP-DF. MAAB acknowledges Marcia Barbosa and Evy Salcedo for useful discussions.

%


\begin{thebibliography}{29}%
\makeatletter
\providecommand \@ifxundefined [1]{%
 \@ifx{#1\undefined}
}%
\providecommand \@ifnum [1]{%
 \ifnum #1\expandafter \@firstoftwo
 \else \expandafter \@secondoftwo
 \fi
}%
\providecommand \@ifx [1]{%
 \ifx #1\expandafter \@firstoftwo
 \else \expandafter \@secondoftwo
 \fi
}%
\providecommand \natexlab [1]{#1}%
\providecommand \enquote  [1]{``#1''}%
\providecommand \bibnamefont  [1]{#1}%
\providecommand \bibfnamefont [1]{#1}%
\providecommand \citenamefont [1]{#1}%
\providecommand \href@noop [0]{\@secondoftwo}%
\providecommand \href [0]{\begingroup \@sanitize@url \@href}%
\providecommand \@href[1]{\@@startlink{#1}\@@href}%
\providecommand \@@href[1]{\endgroup#1\@@endlink}%
\providecommand \@sanitize@url [0]{\catcode `\\12\catcode `\$12\catcode
  `\&12\catcode `\#12\catcode `\^12\catcode `\_12\catcode `\%12\relax}%
\providecommand \@@startlink[1]{}%
\providecommand \@@endlink[0]{}%
\providecommand \url  [0]{\begingroup\@sanitize@url \@url }%
\providecommand \@url [1]{\endgroup\@href {#1}{\urlprefix }}%
\providecommand \urlprefix  [0]{URL }%
\providecommand \Eprint [0]{\href }%
\providecommand \doibase [0]{http://dx.doi.org/}%
\providecommand \selectlanguage [0]{\@gobble}%
\providecommand \bibinfo  [0]{\@secondoftwo}%
\providecommand \bibfield  [0]{\@secondoftwo}%
\providecommand \translation [1]{[#1]}%
\providecommand \BibitemOpen [0]{}%
\providecommand \bibitemStop [0]{}%
\providecommand \bibitemNoStop [0]{.\EOS\space}%
\providecommand \EOS [0]{\spacefactor3000\relax}%
\providecommand \BibitemShut  [1]{\csname bibitem#1\endcsname}%
\let\auto@bib@innerbib\@empty
\bibitem [{\citenamefont {Franks}(2000)}]{Franks:water:matrix}%
  \BibitemOpen
  \bibfield  {author} {\bibinfo {author} {\bibfnamefont {F.}~\bibnamefont
  {Franks}},\ }\href@noop {} {\emph {\bibinfo {title} {{Water: a Matrix for
  life (second edition)}}}}\ (\bibinfo  {publisher} {Royal Society of
  Chemistry},\ \bibinfo {year} {2000})\BibitemShut {NoStop}%
\bibitem [{\citenamefont {Debenedetti}(2003)}]{debenedetti03:review}%
  \BibitemOpen
  \bibfield  {author} {\bibinfo {author} {\bibfnamefont {P.~G.}\ \bibnamefont
  {Debenedetti}},\ }\href@noop {} {\bibfield  {journal} {\bibinfo  {journal}
  {J. Phys.: Cond. Matter}\ }\textbf {\bibinfo {volume} {15}},\ \bibinfo
  {pages} {1669} (\bibinfo {year} {2003})}\BibitemShut {NoStop}%
\bibitem [{\citenamefont {Malenkov}(2009)}]{malenkov2009:jpcm:review}%
  \BibitemOpen
  \bibfield  {author} {\bibinfo {author} {\bibfnamefont {G.}~\bibnamefont
  {Malenkov}},\ }\href@noop {} {\bibfield  {journal} {\bibinfo  {journal} {J.
  Phys.: Condens. Matter}\ }\textbf {\bibinfo {volume} {21}},\ \bibinfo {pages}
  {283101} (\bibinfo {year} {2009})}\BibitemShut {NoStop}%
\bibitem [{\citenamefont {Poole}\ \emph {et~al.}(1992)\citenamefont {Poole},
  \citenamefont {Sciortino}, \citenamefont {Essmann},\ and\ \citenamefont
  {Stanley}}]{poole92:nat}%
  \BibitemOpen
  \bibfield  {author} {\bibinfo {author} {\bibfnamefont {P.~H.}\ \bibnamefont
  {Poole}}, \bibinfo {author} {\bibfnamefont {F.}~\bibnamefont {Sciortino}},
  \bibinfo {author} {\bibfnamefont {U.}~\bibnamefont {Essmann}}, \ and\
  \bibinfo {author} {\bibfnamefont {H.~E.}\ \bibnamefont {Stanley}},\
  }\href@noop {} {\bibfield  {journal} {\bibinfo  {journal} {Nature}\ }\textbf
  {\bibinfo {volume} {360}},\ \bibinfo {pages} {324} (\bibinfo {year}
  {1992})}\BibitemShut {NoStop}%
\bibitem [{\citenamefont {Sastry}\ \emph {et~al.}(1996)\citenamefont {Sastry},
  \citenamefont {Debenedetti}, \citenamefont {Sciortino},\ and\ \citenamefont
  {Stanley}}]{sastry96:pre}%
  \BibitemOpen
  \bibfield  {author} {\bibinfo {author} {\bibfnamefont {S.}~\bibnamefont
  {Sastry}}, \bibinfo {author} {\bibfnamefont {P.~G.}\ \bibnamefont
  {Debenedetti}}, \bibinfo {author} {\bibfnamefont {F.}~\bibnamefont
  {Sciortino}}, \ and\ \bibinfo {author} {\bibfnamefont {H.~E.}\ \bibnamefont
  {Stanley}},\ }\href@noop {} {\bibfield  {journal} {\bibinfo  {journal} {Phys.
  Rev. E}\ }\textbf {\bibinfo {volume} {53}},\ \bibinfo {pages} {6144}
  (\bibinfo {year} {1996})}\BibitemShut {NoStop}%
\bibitem [{\citenamefont {Rebelo}\ \emph {et~al.}(1998)\citenamefont {Rebelo},
  \citenamefont {Debenedetti},\ and\ \citenamefont {Sastry}}]{sastry98:jcp}%
  \BibitemOpen
  \bibfield  {author} {\bibinfo {author} {\bibfnamefont {L.~P.~N.}\
  \bibnamefont {Rebelo}}, \bibinfo {author} {\bibfnamefont {P.~G.}\
  \bibnamefont {Debenedetti}}, \ and\ \bibinfo {author} {\bibfnamefont
  {S.}~\bibnamefont {Sastry}},\ }\href@noop {} {\bibfield  {journal} {\bibinfo
  {journal} {J. Chem. Phys.}\ }\textbf {\bibinfo {volume} {109}},\ \bibinfo
  {pages} {629} (\bibinfo {year} {1998})}\BibitemShut {NoStop}%
\bibitem [{\citenamefont {Franzese}\ \emph {et~al.}(2003)\citenamefont
  {Franzese}, \citenamefont {Marques},\ and\ \citenamefont
  {Stanley}}]{giancarlo03:pre}%
  \BibitemOpen
  \bibfield  {author} {\bibinfo {author} {\bibfnamefont {G.}~\bibnamefont
  {Franzese}}, \bibinfo {author} {\bibfnamefont {M.~I.}\ \bibnamefont
  {Marques}}, \ and\ \bibinfo {author} {\bibfnamefont {H.~E.}\ \bibnamefont
  {Stanley}},\ }\href@noop {} {\bibfield  {journal} {\bibinfo  {journal} {Phys.
  Rev. E}\ }\textbf {\bibinfo {volume} {67}} (\bibinfo {year}
  {2003})}\BibitemShut {NoStop}%
\bibitem [{\citenamefont {Heckmann}\ and\ \citenamefont
  {Drossel}(2013)}]{heckmann2013:jcp}%
  \BibitemOpen
  \bibfield  {author} {\bibinfo {author} {\bibfnamefont {L.}~\bibnamefont
  {Heckmann}}\ and\ \bibinfo {author} {\bibfnamefont {B.}~\bibnamefont
  {Drossel}},\ }\href@noop {} {\bibfield  {journal} {\bibinfo  {journal} {The
  Journal of chemical physics}\ }\textbf {\bibinfo {volume} {138}},\ \bibinfo
  {pages} {234503} (\bibinfo {year} {2013})}\BibitemShut {NoStop}%
\bibitem [{\citenamefont {Barbosa}\ and\ \citenamefont
  {Henriques}(2008)}]{barbosa08}%
  \BibitemOpen
  \bibfield  {author} {\bibinfo {author} {\bibfnamefont {M.~A.~A.}\
  \bibnamefont {Barbosa}}\ and\ \bibinfo {author} {\bibfnamefont {V.~B.}\
  \bibnamefont {Henriques}},\ }\href@noop {} {\bibfield  {journal} {\bibinfo
  {journal} {Phys. Rev. E}\ }\textbf {\bibinfo {volume} {77}},\ \bibinfo
  {pages} {051204} (\bibinfo {year} {2008})}\BibitemShut {NoStop}%
\bibitem [{\citenamefont {{de Oliveira}}\ and\ \citenamefont
  {Barbosa}(2005)}]{marcia05:jcp}%
  \BibitemOpen
  \bibfield  {author} {\bibinfo {author} {\bibfnamefont {A.~B.}\ \bibnamefont
  {{de Oliveira}}}\ and\ \bibinfo {author} {\bibfnamefont {M.~C.}\ \bibnamefont
  {Barbosa}},\ }\href@noop {} {\bibfield  {journal} {\bibinfo  {journal} {J.
  Phys.: Cond. Matter}\ }\textbf {\bibinfo {volume} {17}},\ \bibinfo {pages}
  {399} (\bibinfo {year} {2005})}\BibitemShut {NoStop}%
\bibitem [{\citenamefont {Szortyka}\ \emph {et~al.}(2010)\citenamefont
  {Szortyka}, \citenamefont {Fiore}, \citenamefont {Henriques},\ and\
  \citenamefont {Barbosa}}]{szortyka10:jcp}%
  \BibitemOpen
  \bibfield  {author} {\bibinfo {author} {\bibfnamefont {M.~M.}\ \bibnamefont
  {Szortyka}}, \bibinfo {author} {\bibfnamefont {C.~E.}\ \bibnamefont {Fiore}},
  \bibinfo {author} {\bibfnamefont {V.~B.}\ \bibnamefont {Henriques}}, \ and\
  \bibinfo {author} {\bibfnamefont {M.~C.}\ \bibnamefont {Barbosa}},\
  }\href@noop {} {\bibfield  {journal} {\bibinfo  {journal} {J. Chem. Phys.}\
  }\textbf {\bibinfo {volume} {133}},\ \bibinfo {pages} {104904} (\bibinfo
  {year} {2010})}\BibitemShut {NoStop}%
\bibitem [{\citenamefont {Kumar}\ \emph {et~al.}(2008)\citenamefont {Kumar},
  \citenamefont {Franzese},\ and\ \citenamefont {Stanley}}]{giancarlo08:prl}%
  \BibitemOpen
  \bibfield  {author} {\bibinfo {author} {\bibfnamefont {P.}~\bibnamefont
  {Kumar}}, \bibinfo {author} {\bibfnamefont {G.}~\bibnamefont {Franzese}}, \
  and\ \bibinfo {author} {\bibfnamefont {H.~E.}\ \bibnamefont {Stanley}},\
  }\href@noop {} {\bibfield  {journal} {\bibinfo  {journal} {Phys. Rev. Lett.}\
  }\textbf {\bibinfo {volume} {100}},\ \bibinfo {pages} {105701} (\bibinfo
  {year} {2008})}\BibitemShut {NoStop}%
\bibitem [{\citenamefont {Barbosa}\ \emph {et~al.}(2008)\citenamefont
  {Barbosa}, \citenamefont {Barbosa},\ and\ \citenamefont
  {Oliveira}}]{barbosa11:jcp}%
  \BibitemOpen
  \bibfield  {author} {\bibinfo {author} {\bibfnamefont {M.~A.~A.}\
  \bibnamefont {Barbosa}}, \bibinfo {author} {\bibfnamefont {F.~V.}\
  \bibnamefont {Barbosa}}, \ and\ \bibinfo {author} {\bibfnamefont {F.~A.}\
  \bibnamefont {Oliveira}},\ }\href@noop {} {\bibfield  {journal} {\bibinfo
  {journal} {J. Chem. Phys.}\ }\textbf {\bibinfo {volume} {134}},\ \bibinfo
  {pages} {024511} (\bibinfo {year} {2008})}\BibitemShut {NoStop}%
\bibitem [{\citenamefont {Barbosa}\ \emph {et~al.}(2013)\citenamefont
  {Barbosa}, \citenamefont {Salcedo},\ and\ \citenamefont
  {Barbosa}}]{barbosa13:pre}%
  \BibitemOpen
  \bibfield  {author} {\bibinfo {author} {\bibfnamefont {M.~A.~A.}\
  \bibnamefont {Barbosa}}, \bibinfo {author} {\bibfnamefont {E.}~\bibnamefont
  {Salcedo}}, \ and\ \bibinfo {author} {\bibfnamefont {M.}~\bibnamefont
  {Barbosa}},\ }\href@noop {} {\bibfield  {journal} {\bibinfo  {journal} {Phys.
  Rev. E}\ }\textbf {\bibinfo {volume} {87}},\ \bibinfo {pages} {032303}
  (\bibinfo {year} {2013})}\BibitemShut {NoStop}%
\bibitem [{\citenamefont {Errington}\ and\ \citenamefont
  {Debenedetti}(2001)}]{debenedetti01:nat}%
  \BibitemOpen
  \bibfield  {author} {\bibinfo {author} {\bibfnamefont {J.~R.}\ \bibnamefont
  {Errington}}\ and\ \bibinfo {author} {\bibfnamefont {P.~G.}\ \bibnamefont
  {Debenedetti}},\ }\href@noop {} {\bibfield  {journal} {\bibinfo  {journal}
  {Nature}\ }\textbf {\bibinfo {volume} {409}},\ \bibinfo {pages} {318}
  (\bibinfo {year} {2001})}\BibitemShut {NoStop}%
\bibitem [{Note1()}]{Note1}%
  \BibitemOpen
  \bibinfo {note} {Notation for this phase has changed from bonded fluid to
  softened fluid, since our previous work, because there are no~\protect
  \textit {hydrogen bonded} in the current context.}\BibitemShut {Stop}%
\bibitem [{\citenamefont {Sadr-Lahijany}\ \emph {et~al.}(1999)\citenamefont
  {Sadr-Lahijany}, \citenamefont {Scala}, \citenamefont {Buldyrev},\ and\
  \citenamefont {Stanley}}]{stanley99:pre}%
  \BibitemOpen
  \bibfield  {author} {\bibinfo {author} {\bibfnamefont {M.~R.}\ \bibnamefont
  {Sadr-Lahijany}}, \bibinfo {author} {\bibfnamefont {A.}~\bibnamefont
  {Scala}}, \bibinfo {author} {\bibfnamefont {S.~V.}\ \bibnamefont {Buldyrev}},
  \ and\ \bibinfo {author} {\bibfnamefont {H.~E.}\ \bibnamefont {Stanley}},\
  }\href@noop {} {\bibfield  {journal} {\bibinfo  {journal} {Phys. Rev. E}\
  }\textbf {\bibinfo {volume} {60}},\ \bibinfo {pages} {6714} (\bibinfo {year}
  {1999})}\BibitemShut {NoStop}%
\bibitem [{\citenamefont {Barbosa}\ and\ \citenamefont
  {Widom}(2010)}]{barbosa10:jcp}%
  \BibitemOpen
  \bibfield  {author} {\bibinfo {author} {\bibfnamefont {M.~A.~A.}\
  \bibnamefont {Barbosa}}\ and\ \bibinfo {author} {\bibfnamefont
  {B.}~\bibnamefont {Widom}},\ }\href@noop {} {\bibfield  {journal} {\bibinfo
  {journal} {J. Chem. Phys.}\ }\textbf {\bibinfo {volume} {132}},\ \bibinfo
  {pages} {214506} (\bibinfo {year} {2010})}\BibitemShut {NoStop}%
\bibitem [{\citenamefont {Salinas}(2001)}]{salinas:introduction}%
  \BibitemOpen
  \bibfield  {author} {\bibinfo {author} {\bibfnamefont {S.~R.~A.}\
  \bibnamefont {Salinas}},\ }\href@noop {} {\emph {\bibinfo {title}
  {{Introduction to statistical physics}}}}\ (\bibinfo  {publisher}
  {Springer-Verlag},\ \bibinfo {address} {New York},\ \bibinfo {year} {2001})\
  \bibinfo {note} {p. 47}\BibitemShut {NoStop}%
\bibitem [{\citenamefont {Binder}\ and\ \citenamefont
  {Heermann}(1988)}]{binder}%
  \BibitemOpen
  \bibfield  {author} {\bibinfo {author} {\bibfnamefont {K.}~\bibnamefont
  {Binder}}\ and\ \bibinfo {author} {\bibfnamefont {D.~W.}\ \bibnamefont
  {Heermann}},\ }\href@noop {} {\emph {\bibinfo {title} {{Monte {C}arlo
  simulation in statistical physics}}}}\ (\bibinfo  {publisher}
  {Springer-Verlag},\ \bibinfo {year} {1988})\BibitemShut {NoStop}%
\bibitem [{com()}]{comment2}%
  \BibitemOpen
  \href@noop {} {}\bibinfo {note} {As noted by one of the referees, a Monte
  Carlo simulation is a method for sampling the statistical ensemble with an
  ``artificial'' dynamics. Nevertheless, since the classical lattice gas model
  investigated in this work does not evolve in time through Newton's laws or
  Schrodinger equation, it is necessary to define the system's dynamics by
  choosing an stochastic sampling method and a set of allowed movements. As
  usual, dynamics must converge to equilibrium at sufficiently large
  times.}\BibitemShut {Stop}%
\bibitem [{\citenamefont {Girardi}\ \emph {et~al.}(2007)\citenamefont
  {Girardi}, \citenamefont {Szortyka},\ and\ \citenamefont
  {Barbosa}}]{marcia07:pa}%
  \BibitemOpen
  \bibfield  {author} {\bibinfo {author} {\bibfnamefont {M.}~\bibnamefont
  {Girardi}}, \bibinfo {author} {\bibfnamefont {M.}~\bibnamefont {Szortyka}}, \
  and\ \bibinfo {author} {\bibfnamefont {M.~C.}\ \bibnamefont {Barbosa}},\
  }\href@noop {} {\bibfield  {journal} {\bibinfo  {journal} {Physica A}\
  }\textbf {\bibinfo {volume} {386}},\ \bibinfo {pages} {692} (\bibinfo {year}
  {2007})}\BibitemShut {NoStop}%
\bibitem [{\citenamefont {Szortyka}\ and\ \citenamefont
  {Barbosa}(2007)}]{marcia07:pa2}%
  \BibitemOpen
  \bibfield  {author} {\bibinfo {author} {\bibfnamefont {M.}~\bibnamefont
  {Szortyka}}\ and\ \bibinfo {author} {\bibfnamefont {M.~C.}\ \bibnamefont
  {Barbosa}},\ }\href@noop {} {\bibfield  {journal} {\bibinfo  {journal}
  {Physica A}\ }\textbf {\bibinfo {volume} {380}},\ \bibinfo {pages} {27}
  (\bibinfo {year} {2007})}\BibitemShut {NoStop}%
\bibitem [{\citenamefont {Angell}\ \emph {et~al.}(1976)\citenamefont {Angell},
  \citenamefont {Finch},\ and\ \citenamefont {Bach}}]{angell:jcp76}%
  \BibitemOpen
  \bibfield  {author} {\bibinfo {author} {\bibfnamefont {C.~A.}\ \bibnamefont
  {Angell}}, \bibinfo {author} {\bibfnamefont {E.~D.}\ \bibnamefont {Finch}}, \
  and\ \bibinfo {author} {\bibfnamefont {P.}~\bibnamefont {Bach}},\ }\href@noop
  {} {\bibfield  {journal} {\bibinfo  {journal} {J. Chem. Phys.}\ }\textbf
  {\bibinfo {volume} {65}},\ \bibinfo {pages} {3063} (\bibinfo {year}
  {1976})}\BibitemShut {NoStop}%
\bibitem [{\citenamefont {Hu}\ and\ \citenamefont
  {Izmailian}(1998)}]{hu98:pre}%
  \BibitemOpen
  \bibfield  {author} {\bibinfo {author} {\bibfnamefont {C.-K.}\ \bibnamefont
  {Hu}}\ and\ \bibinfo {author} {\bibfnamefont {N.~S.}\ \bibnamefont
  {Izmailian}},\ }\href@noop {} {\bibfield  {journal} {\bibinfo  {journal}
  {Phys. Rev. E}\ }\textbf {\bibinfo {volume} {58}},\ \bibinfo {pages} {1644}
  (\bibinfo {year} {1998})}\BibitemShut {NoStop}%
\bibitem [{Note2()}]{Note2}%
  \BibitemOpen
  \bibinfo {note} {While crossing the neighborhood of the ground state phase
  transitions, in the $P$ vs. $T$ phase diagram, it is possible to observe the
  discontinuity in extensive variables (\protect \textit {e.g.} volume) and the
  diverging behavior on response functions (\protect \textit {e.g.} thermal
  expansion coefficient)}\BibitemShut {NoStop}%
\bibitem [{\citenamefont {S{\'a}nchez}\ \emph {et~al.}(1987)\citenamefont
  {S{\'a}nchez}, \citenamefont {Costas},\ and\ \citenamefont
  {Varea}}]{sanches87}%
  \BibitemOpen
  \bibfield  {author} {\bibinfo {author} {\bibfnamefont {G.}~\bibnamefont
  {S{\'a}nchez}}, \bibinfo {author} {\bibfnamefont {M.~E.}\ \bibnamefont
  {Costas}}, \ and\ \bibinfo {author} {\bibfnamefont {C.}~\bibnamefont
  {Varea}},\ }\href@noop {} {\bibfield  {journal} {\bibinfo  {journal} {J.
  Chem. Phys.}\ }\textbf {\bibinfo {volume} {86}},\ \bibinfo {pages} {2966}
  (\bibinfo {year} {1987})}\BibitemShut {NoStop}%
\bibitem [{\citenamefont {Russo}\ and\ \citenamefont
  {Tanaka}(2014)}]{tanaka2014:nature}%
  \BibitemOpen
  \bibfield  {author} {\bibinfo {author} {\bibfnamefont {J.}~\bibnamefont
  {Russo}}\ and\ \bibinfo {author} {\bibfnamefont {H.}~\bibnamefont {Tanaka}},\
  }\href@noop {} {\bibfield  {journal} {\bibinfo  {journal} {Nat. Commun.}\
  }\textbf {\bibinfo {volume} {5}},\ \bibinfo {pages} {3556} (\bibinfo {year}
  {2014})}\BibitemShut {NoStop}%
\bibitem [{\citenamefont {Cuthbertson}\ and\ \citenamefont
  {Poole}(2011)}]{poole11:prl}%
  \BibitemOpen
  \bibfield  {author} {\bibinfo {author} {\bibfnamefont {M.~J.}\ \bibnamefont
  {Cuthbertson}}\ and\ \bibinfo {author} {\bibfnamefont {P.~H.}\ \bibnamefont
  {Poole}},\ }\href@noop {} {\bibfield  {journal} {\bibinfo  {journal} {Phys.
  Rev. Lett.}\ }\textbf {\bibinfo {volume} {106}},\ \bibinfo {pages} {115706}
  (\bibinfo {year} {2011})}\BibitemShut {NoStop}%
\end{thebibliography}

\end{document}